\documentclass[aps,amsfonts,prl,twocolumn,showpacs]{revtex4-1}
\usepackage{epsfig,amsmath,amssymb,bm,epsf,graphics,psfrag,verbatim,subfigure}

\def\beq{\begin{equation}}
\def\eeq{\end{equation}}
\def\bea{\begin{eqnarray}}
\def\eea{\end{eqnarray}}

\begin{document}

\title{Universal phase structure of dilute Bose gases with Rashba spin-orbit coupling}

\author{Sarang Gopalakrishnan$^{1,2}$,  Austen Lamacraft$^{2,3}$, and Paul M. Goldbart$^{4}$}
\affiliation{
$^1$Department of Physics, University of Illinois at Urbana-Champaign, Urbana, Illinois 61801}

\affiliation{$^2$Kavli Institute for Theoretical Physics, 
University of California, Santa Barbara, California 93106}

\affiliation{$^3$Department of Physics, University of Virginia, Charlottesville, Virginia 22904}

\affiliation{$^4$School of Physics, Georgia Institute of Technology, Atlanta, Georgia 30332}

\date{June 13, 2011}

\begin{abstract}
A Bose gas subject to a light-induced Rashba spin-orbit coupling possesses a dispersion minimum on a circle in momentum space; we show that kinematic constraints due to this dispersion cause interactions to renormalize to universal, angle-dependent values that govern the phase structure in the dilute-gas limit. We find that, regardless of microscopic interactions, (a)~the ground state involves condensation at two opposite momenta, and is, in finite systems, a fragmented condensate; and (b)~there is a nonzero-temperature instability toward the condensation of pairs of bosons. We discuss how our results can be reconciled with the qualitatively different mean-field phase diagram, which is appropriate for dense gases.
\end{abstract}

\maketitle

The advent of ultracold gases has vastly increased the range of physically realizable many-body bosonic systems, enabling the exploration of quantum-degenerate Bose gases possessing tunable interactions and band structure as well as internal degrees of freedom. 
Among such systems, those of particular interest involve single-particle Hamiltonians having degenerate ground states related by symmetries. Bose-Einstein condensation (BEC)---i.e., the macroscopic occupation of a \textit{particular} single-particle state---typically entails breaking these symmetries; hence the order parameter space and defects of such BECs are richer than those of conventional BECs. 
For instance, spin-1 BECs~\cite{ueda} support fractionally quantized vortices, and in this sense resemble exotic \textit{fermionic} condensates such as triplet superconductors. 

Just as these exotic defects stem from broken \textit{internal} symmetries, those of the Fulde-Ferrell-Larkin-Ovchinnikov (FFLO) state, such as vortex-dislocation bound states~\cite{radz}, stem from its broken translational and rotational symmetries. 
The present work addresses purely bosonic analogs of the FFLO states, in which the degenerate single-particle ground states have distinct \textit{spatial} wavefunctions. 
In particular, we consider the case in which the single-particle Hamiltonian possesses a dispersion minimum on a circle in momentum space, so that BEC occurs at one or more nonzero momenta. Our work is motivated by a recently proposed realization of such a Hamiltonian, viz.~a spin-$\frac{1}{2}$ Bose gas subject to a light-induced Rashba spin-orbit coupling~\cite{campbell}. Simpler forms of spin-orbit coupling, having multiple discrete minima, have been experimentally demonstrated~\cite{spielman}. An alternative approach to realizing a circular dispersion minimum would be to load the atoms into the excited band of an optical lattice; in this case, too, multiple discrete minima have been realized~\cite{hemmerich}, and under appropriate conditions (e.g., ``SE-even faulted''\ stackings of bilayer honeycomb lattices~\cite{mele}) continuous minima are realizable. 

\begin{figure}[b]
	\centering
		\includegraphics{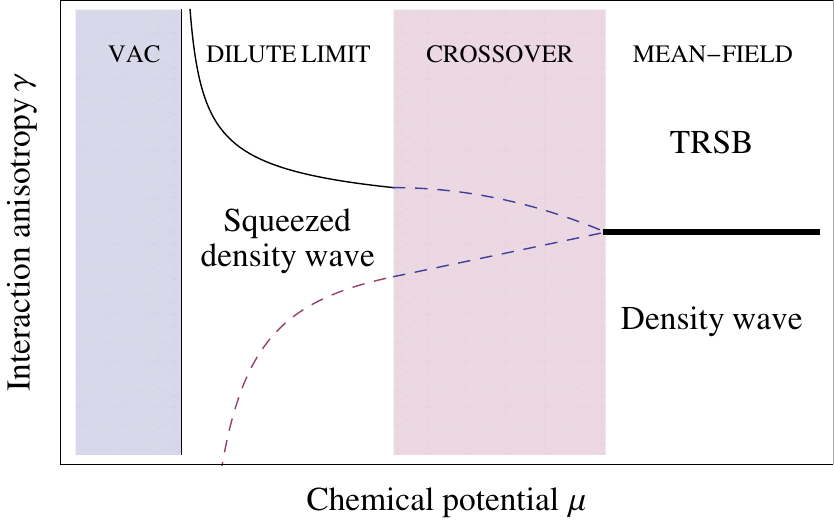}
	\vspace{-12pt}
	\caption{Zero-temperature phase diagram as a function of the interaction anisotropy $\gamma \equiv c_2/c_0$ and the chemical potential $\mu$, showing the phases and transitions discussed in the main text. The (dashed) phase boundaries in the crossover region are schematic; the bold line indicates a first-order transition predicted by mean-field theory~\cite{zhai}.}
	\label{fig:fig2}
\end{figure}

Spin-orbit coupled BECs were originally addressed in Refs.~\cite{galitski, mondragon} as examples of unconventional condensation; it was argued in Ref.~\cite{mondragon} that, for a pure Rashba coupling and \textit{isotropic} interactions, a fragmented condensate should form. More recently, the case of the Rashba-coupled BEC was treated using mean-field theory~\cite{zhai} and incorporating Gaussian fluctuations~\cite{zhai2}; related systems have been studied in Refs.~\cite{generic}. In general, two phases have been found, depending on the spin-dependence of interactions: a time-reversal-symmetry-breaking (TRSB) state and a density-wave state. In the present work, we describe how interaction-renormalization effects \textit{qualitatively} change the phase diagram at low densities (see Fig.~\ref{fig:fig2}), destabilizing the TRSB state and giving rise to a number-squeezed (and, in finite systems, ``fragmented'') limit of the density-wave state. These changes are due to the strong, \textit{emergent} angle-dependence of renormalized interactions; such angle-dependent renormalizations are generic in systems whose low-energy modes occur around momentum-space surfaces, e.g., Fermi liquids~\cite{shankar}. Our results, while consistent with those of Ref.~\cite{mondragon} in the special case of isotropic interactions, hold more generally for \textit{any} interactions that are repulsive in all angular-momentum channels. 

Our primary results are as follows. At zero temperature, we find---exploiting the properties of a quantum critical point introduced in Ref.~\cite{yang:sachdev}---that the renormalized interactions for a dilute gas \textit{universally} favor a state in which the BEC forms at a pair of opposite momenta. For \textit{finite}, weakly trapped systems, fragmented BEC is energetically favored over simple BEC at either a single momentum or a coherent momentum superposition such as a density wave. In the thermodynamic limit, the fragmented BEC, though favored over a \textit{coherent} superposition, becomes degenerate with \textit{squeezed} states that break translational symmetry. The resulting ground-state energy per particle scales unusually with the density $n$, i.e., as $n^{4/3}$; note that this scaling is the same as that of the ``extremely anisotropic Wigner crystal''~\cite{berg}, which in fact approaches the fragmented state in the zero-density limit. At nonzero temperature, we argue using renormalization-group (RG) methods that the leading instability is toward condensation of boson \textit{pairs}, and estimate the condensation temperature. 

\textit{Model}. We begin with the following effective model of a $d$-dimensional Bose gas having a circular dispersion minimum: 

\bea
H & = & \int d^d k \, \Psi^\dagger(\mathbf{k})\! \left[ -\mu + \frac{1}{2M}\{(|\mathbf{k}_{2D}| - k_0)^2 + k_\perp^2\} \right]\! \Psi(\mathbf{k}) \nonumber \\
& & + \int \prod_{i = 1}^{4} d^d k_i\, U(\{\mathbf{k}_i\}) \Psi^\dagger(\mathbf{k}_1) \Psi^\dagger(\mathbf{k}_2) \Psi(\mathbf{k}_3) \nonumber \\
&& \qquad \qquad \times \Psi(\mathbf{k}_1 + \mathbf{k}_2 - \mathbf{k}_3),
\eea
where $\Psi(\mathbf{k}_i)$ are Bose fields of momentum $\mathbf{k}_i$; $\mathbf{k}_{2D} \equiv (k_x,k_y)$; $k_\perp$ encodes all other momentum components; $U$ is a possibly momentum-dependent interaction; and we have set $\hbar = 1$. For Rashba-coupled bosons, \textit{spin}-dependent interactions in the microscopic model imply momentum-dependent interactions because, for modes near $k_0$, the spin is locked to the momentum. We shall first consider the universal properties of the general Hamiltonian $H$, and then relate these to the phases of the specific microscopic model considered in Ref.~\cite{zhai}.  
We focus primarily on the 2D case, in which $k_\perp = 0$, and then touch on the (similar) 3D case. 


We assume that energies associated with temperature $T$, chemical potential $\mu$, system size, etc. are smaller than the spin-orbit coupling scale $k_0^2/2M$. Typical values of $k^{-1}_0$ are on the order of an optical wavelength~\cite{spielman}, which is exceeded by the interparticle spacing in many experiments ($k_0^2/2M$ cannot be \emph{smaller} than these scales if spin-orbit coupling is to play a significant role).

As we are concerned with the low-energy limit, it is convenient to study only the degrees of freedom in a momentum shell of thickness $2\Lambda$ centered on the dispersion minimum, giving rise to an energy scale $\Omega_{\Lambda}\equiv\Lambda^{2}/2M$ intermediate between $k_{0}/2M$ and the low-energy scales $\mu$ and $T$. Integrating out degrees of freedom with energies $\geq \Omega_\Lambda$ generates effective interactions for modes with energies $\leq \Omega_\Lambda$; as we discuss below, these interactions are further renormalized, and (for energies $\ll \Omega_\Lambda$, take on universal values that are independent of $\Lambda$. A careful treatment of the high-energy renormalization, including the other Rashba bands, appears in a recent preprint \cite{ozawa} and confirms this picture.

\begin{figure}
	\centering
		\includegraphics{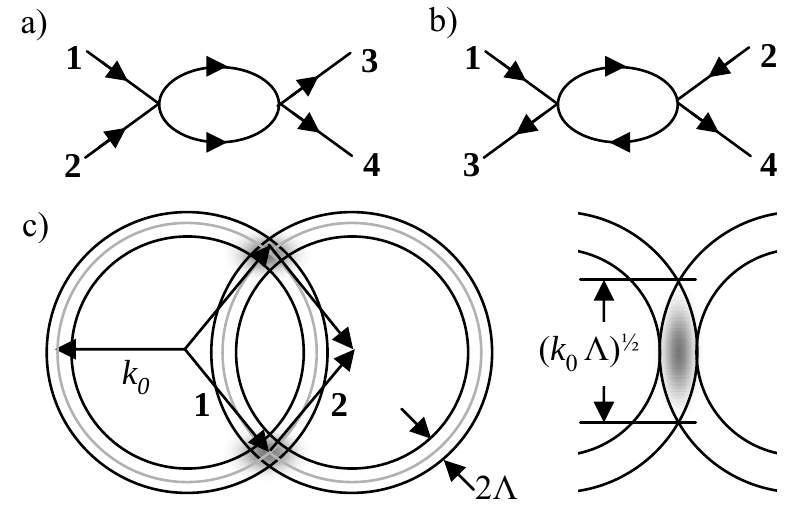}
	\vspace{-12pt}
	\caption{(a) Loop correction in the particle-particle channel, which governs the hierarchy of couplings at the QCP. (b) Loop correction in the particle-hole channel. These corrections vanish at $T = 0$. (c) Kinematic constraints due to the dispersion structure: left, case of $\theta \neq 0$: outgoing momenta are constrained to lie in the shaded region, of area $\sim \Lambda^2$; right, case of $\theta = 0$, for which the shaded region's area scales as $\Lambda \sqrt{k_0 \Lambda}$.}
	\label{fig:fig1}
	\vspace{-6pt}
\end{figure}

\textit{Quantum critical point}. 
The model described by $H$ has a quantum critical point (QCP) at $\mu = 0$, corresponding to the phase transition from the empty vacuum to a BEC. This QCP was analyzed in Ref.~\cite{yang:sachdev} \textit{for fermions}, but the analysis extends straightforwardly to bosons. Given that $\Lambda \ll k_0$, kinematics constrains the resulting form of the effective interaction vertices within the momentum shell (i.e., those for which all four momenta satisfy $||\mathbf{k}_i| - k_0| \leq \Lambda$) to lie in the following channels: (i)~forward scattering processes, which involve momentum transfer $\alt \Lambda$ [denoted $F_{\Omega_{\Lambda}}(\theta)$ where $\theta$ is the angle between the incoming momenta]; and (ii)~``Cooper-channel''\ processes, in which incoming momenta are almost equal and opposite [denoted $V_{\Omega_{\Lambda}}(\theta)$ where $\theta$ is the angle between incoming and outgoing momentum pairs~(see, e.g., Ref.~\cite{shankar})]. These channels renormalize differently: owing to the non-polarizability of the vacuum~\cite{sachdev:book}, all renormalizations are due to the repeated scattering processes shown in Fig.~\ref{fig:fig1}a, which have different amplitudes in the forward-scattering and Cooper channels. For forward scattering, intermediate momenta are constrained to lie in the regions shaded in Fig.~\ref{fig:fig1}c, whereas in the Cooper channel intermediate momenta run over the entire circle of radius $k_0$. 

The outcome of renormalization depends on the sign of the microscopic interactions. \textit{Any} attractive interactions lead to an instability in the Cooper channel~\cite{yang:sachdev}, and thereby to bound states; this case is not expected to yield universal behavior. If, on the other hand, the initial interactions are \textit{all} repulsive, one arrives at the following expressions for the renormalized interactions, for incoming frequencies of order $\Omega \leq \Omega_{\Lambda}$ (see Ref.~\cite{yang:sachdev}):

\begin{subequations}
\bea
F_\Omega(\theta) & = & \frac{F_{\Omega_{\Lambda}}(\theta)}{1 + [M F_{\Omega_{\Lambda}}(\theta) / (2\pi\sin \theta)] \ln\left(\frac{\Omega_{\Lambda}}{\Omega}\right)}, \\
F_\Omega(\theta\! =\! 0) & = & \frac{F_{\Omega_{\Lambda}}(0)}{1 + M F_{\Omega_{\Lambda}}(0) \sqrt{k_0 / \sqrt{M \Omega}} \,f_1 \!\left(\frac{\Omega_{\Lambda}}{\Omega}\right)},\\
V_\Omega(m) & = & \frac{V_\infty(m)}{1 + M V_{\Omega_{\Lambda}}(m) \left(k_0 / \sqrt{M \Omega}\right) f_2\!\left(\frac{\Omega_{\Lambda}}{\Omega} \right)},
\eea
\end{subequations}
where $f_1(x)$ and $f_2(x)$ are scaling functions that are of order unity as $x \rightarrow \infty$ and approach zero as $x \rightarrow 0$; and $V(m) \equiv \int_0^{2\pi} V(\theta) e^{im\theta} d\theta$. Subscripts $\Omega$ denote the incoming frequencies.
Thus the low-energy (i.e., $\Omega/\Omega_{\Lambda}\to 0$) values of all couplings are ``universal,'' i.e.,~independent of their microscopic values. Note that $F(\theta = \pi) = \sum\nolimits_{m \, \, \mathrm{even}} V(m)$. Thus, given that $\Lambda/k_0 \ll 1$, the couplings assume the following hierarchy: $V_\Omega(m) \sim F_\Omega(\theta = \pi) \ll F_\Omega(\theta = 0) \ll F_\Omega(\theta \neq 0, \pi)$. Hence, interactions between particles at opposite momenta are negligible compared with other interactions~\cite{fn1}.

\textit{Case of T = 0}.
We now turn from the QCP to phases in its vicinity. Suppose that the system is sufficiently dilute that when renormalization is cut off at a scale set by the chemical potential $\mu$, the interactions are deep in the universal scaling regime.
%
%
Then the interaction Hamiltonian is given by $H \sim \sum\nolimits_{\theta, \theta'} F(\theta - \theta') n_\theta n_{\theta'}$, where $n_\theta$ denotes the boson density at a momentum of magnitude $k_0$ and direction $\theta$. The hierarchy of universal coupling constants implies that $H$ is minimized by a ``fragmented'' state, having precisely $N/2$ bosons at some $\theta$, and $N/2$ at $\theta + \pi$~\cite{fn2}. Fragmentation is favored owing to a momentum-space analog of Coulomb blockade (cf. Sec. 2.6 of Ref.~\cite{leggett}): bosons with opposite momenta do not interact with one another to leading order in $\sqrt{\Lambda/k_0}$, whereas those at non-opposite momenta do interact.  

In more quantitative terms we can deduce the ground-state energy from the relation~\cite{fh} $\mu = (n/2) F_\mu(\theta = 0) \simeq (n/M) (\mu M / k_0^2)^{1/4}$, giving

\beq
E(N) \simeq \sum_{\sigma = \pm} \frac{\hbar^2 N_\sigma^{7/3}}{M \mathcal{A}^{4/3} k_0^{2/3}}.
\eeq
where $\mathcal{A}$ is the system area, and $N_\pm$ respectively denote the number of particles at $\theta$ and $\theta + \pi$. Note that this expression is \textit{universal}, i.e., independent of the microscopic interaction strengths, and its unusual scaling is a consequence of the renormalization discussed above. As $E(N)$ is minimized when $N_+ = N_- = N/2$, the ground state for finite $N$ is fragmented. 
Such a fragmented state can be understood as a density wave of wavevector $k_0$ along the direction $\theta$ with a randomly varying phase (analogous, e.g., to interfering independent condensates~\cite{matthews}). 

Fragmented states are typically unstable relative to simple condensates (i.e., those having a fixed phase relation) because spatial inhomogeneities tend to phase-lock the fragments~\cite{leggett}. In the present case, a phase-locked, coherent superposition would involve fluctuations of order $\sqrt{N}$ in $N_\pm$, and hence cost an energy of order unity relative to the fragmented state even in the thermodynamic limit. Thus, a few  scattering sites cannot overcome the tendency toward fragmentation. Similarly, a \textit{weak} harmonic potential (i.e., of characteristic length much larger than the interparticle spacing) would \textit{not} stabilize a coherent superposition relative to a fragmented state, even in the thermodynamic limit, provided that---according to the standard prescription---the trap frequency $\omega \rightarrow 0$ so as to keep $N \omega^2$ constant. This is because the typical matrix element between $\pm k$ due to the trap is of order $\exp(-2k_0^2 N)$, which rapidly decreases as $N \rightarrow \infty$.

Although a coherent superposition is disfavored in the thermodynamic limit, the energy cost of number fluctuations of order \textit{unity} vanishes as $1/N$. Thus, the thermodynamic ground state (e.g., in a trap) is likely to be a squeezed state with small but nonvanishing phase variance, as opposed to the fragmented state, in which the phase is entirely random. This observation extends to translation-invariant systems, which should therefore exhibit spontaneously broken translational invariance in the thermodynamic limit. 

\textit{Implications for $T = 0$ phase diagram}. The dilute-limit phase diagram is \textit{simpler} than that obtained from mean-field theory: it predicts that BEC occurs at two momenta regardless of microscopic interactions, provided these are repulsive. 
By contrast, mean-field theory~\cite{zhai} predicts a time-reversal-symmetry breaking (TRSB) state or a density-wave state, depending on microscopic interactions. We now give an account of the crossover between universal and mean-field regimes, and estimate the minimum densities required for mean-field results to apply.

The dilute-gas results apply when, upon renormalization, the pertinent interactions have already achieved their universal forms at a length-scale shorter than the interparticle spacing; thus, a TRSB state is disfavored if $F_\mu(\pi) \leq F_\mu(0)$, regardless of whether the (larger) $F_\mu(\theta \neq 0, \pi)$ couplings have approached their universal values. Note that $F_\Lambda(\theta = 0, \pi)$ are related to the parameters $c_0$ and $c_2$ in Ref.~\cite{zhai} as follows: $F_\Lambda(\pi) / F_\Lambda(0) \approx 1 + c_2 / c_0$. [These relations, and similar ones for other couplings, can be derived as outlined following Eq.~(3) in Ref.~\cite{zhai}. Provided $c_2 \alt c_0$, all microscopic couplings are of comparable magnitude.] Therefore, in terms of $c_0$ and $c_2$, the TRSB state is favored only if

\beq
\frac{c_0}{1 + \frac{M}{2\pi} c_0 \sqrt{k_0 / n^{1/2}}} < \frac{c_0 + c_2}{1 + q M(c_0 + c_2) (k_0/n^{1/2})}.
\eeq
where $q$ is a constant of order unity. 

Note that, in addition to the TRSB phase, the Hamiltonian of Ref.~\cite{zhai} also exhibits a regime in which a coherent superposition is \textit{lower} in energy than the fragmented state, owing to terms of the form $\psi^\dagger_{2\mathbf{k}_0} \psi^\dagger_{-\mathbf{k}_0} \psi_{\mathbf{k}_0} \psi_0$, which involve momenta of order $2 k_0$ and thus do not appear in $H$. 

These considerations lead us to the phase diagram shown in Fig.~\ref{fig:fig2}, in which there is no \textit{direct} transition from the vacuum to the TRSB state. The transition from the vacuum to the density-wave state is unusual in being a \textit{continuous} transition (\textit{known} to be continuous as the properties of the QCP are understood exactly~\cite{yang:sachdev}) at which both rotational and translational symmetry are broken. As a general rule (see, e.g. Refs.~\cite{radz, brazovskii}) transitions that break rotational \textit{and }translational symmetry are first-order. For densities $\agt \Lambda^2$, at which the renormalization effects discussed in the present work are not present, mean-field simulations show evidence of metastability~\cite{zhai}; this would suggest a first-order transition between the density-wave and TRSB states.

\textit{Case of $T > 0$.} Following standard treatments of the dilute Bose gas~\cite{fh} we assume that $T \gg \mu$. Beyond the momentum scale $\Lambda_T \equiv 1/\sqrt{2M T}$, the physics is captured by a classical free-energy functional of the form

\beq
S  =  \int d^d k \left[ - \mu + (|\mathbf{k} - k_0)^2 \right] |\psi_{\mathbf{k}}|^2 + S_4, 
\eeq
where $S_4$ denotes the set of angle- and channel-dependent couplings, and we have set $2M = 1$. $S$ is a complex-field version of Brazovskii's model~\cite{brazovskii} (the relevance of Brazovskii's model was previously suggested in Ref.~\cite{galitski}). The initial values for the couplings in $S_4$ are the renormalized interactions at a scale $\Omega = T$. 
At scales $\alt \Lambda_T$, the vacuum is nontrivial, owing to the presence of thermal particles; hence, all couplings are renormalized by the particle-hole channel [Fig.~\ref{fig:fig1}b]. It is convenient to expand $F$ as well as $V$ in terms of angular momenta. One can then implement the momentum-shell RG procedure described in Ref.~\cite{hohswift}, by integrating out modes satisfying $\Lambda_T (1 - dl) < |k - k_0| < \Lambda_T$ and rescaling $k \rightarrow (1 + dl) k, \psi \rightarrow [1 - (3/2) dl] \psi$, and $\mu \rightarrow \mu/\Lambda_T^2$. The couplings transform as follows (ignoring the flow of $\mu$):

\begin{subequations}
\bea
\frac{dF_l(m)}{dl} & = & 3 F_l(m) - \frac{A \, F_l^2(m)}{(1 - \mu_l)^2} - \frac{A}{2} \frac{\sum\nolimits_m V_l^2(m)}{(1 - \mu_l)^2}, \\
\frac{dV_l(m)}{dl} & = & 3 V_l(m) - \frac{A \, V_l^2(m)}{2 (1 - \mu_l)^2} - A \frac{\sum\nolimits_m F_l^2(m)}{(1 - \mu_l)^2},
\eea
\end{subequations}
where $A \equiv 2\pi k_0/\Lambda_T$. If the coupling constants at $\Lambda_T$ are in the universal regime, one can use the fact that $V \ll F$ to drop terms of order $V^2$. The last term in the flow equations drives all \textit{even} $V(m)$ (which are initially near zero) to negative values at some $\Lambda_2 = \Lambda_T (1 - o(\Lambda_T/k_0))$, triggering a runaway growth of the even-parity $V(m)$ couplings. Such runaway growth is associated with a pairing instability, which should in principle occur simultaneously in all even-$m$ channels. (However, as noted in Ref.~\cite{mondragon}, the confining trap acts as a kinetic energy term of the form $\nabla^2_\theta$, which penalizes higher-$m$ channels.)

The pair-condensation temperature can be estimated by observing that arbitrarily weak attractive interactions in the Cooper channel give rise to pairs~\cite{marsiglio} whose binding energy is
$\Delta \sim M V^2 k_0^2$. 
Pairing is favored for $\Delta \geq T$. As $T/E_0 \simeq (\Lambda_1/k_0)^2 \ll (\Lambda/k_0)^2 \ll V \simeq \ln(\Lambda/\Lambda_1)$, one expects pairs to be tightly bound at length-scales comparable to $1/\Lambda_T$; 
at longer distances they can be treated as nonoverlapping. The system is thus a dilute gas of pairs, which condense at a temperature given implicitly~\cite{fh} by $T_c \approx (\hbar^2 n / 4m) \times 1/\ln \ln (n a^2)$, where $a$ is an effective dimer-dimer scattering range, which is of order $\Lambda$. 

Pairing would be straightforward to detect experimentally via radio-frequency spectroscopy, which should reveal a peak corresponding to the pair binding energy; moreover, a pair condensate would support half-quantized vortices detectable via rotation. 

\textit{3D case}. For this case, $k_z \equiv k_\perp$ in Eq.~(1); thus, the dispersion minimum is \textit{circular} rather than spherical, and imposes the same kinematic constraints as in 2D. The 2D analysis thus generalizes readily; the chief difference is that the forward-scattering couplings in 3D renormalize to nonuniversal T-matrices rather than universal values, and the ground-state energy thus depends on microscopic couplings. However, Cooper-channel couplings approach the following universal expression as $\Omega /\Omega_{\Lambda} \rightarrow 0$:

\beq
V_\Omega \sim 1/ \left[ k_0 M \ln(\Omega_{\Lambda} / \Omega) \right].
\eeq
Hence, as in 2D, $\sum\nolimits_m V(m) \sim F(\theta = \pi) \ll F(\theta \neq \pi)$ at low energies. It follows that the dilute-limit ground state universally preserves time-reversal symmetry. This qualitative resemblance to 2D extends to the $T > 0$ case, in which the free-energy functional---in this case, the variant of Brazovskii's model having two transverse dimensions discussed in Ref.~\cite{us}---develops a pairing instability as in 2D. As the Cooper-channel couplings approach universal values more slowly, however, the conditions for the dilute limit to obtain are more stringent in 3D than in 2D. 

\textit{Acknowledgments}. S.G. is indebted to H. Zhai, C. Xu, and M. Kindermann for helpful discussions. This work was supported by DOE DE-FG02-07ER46453 (S.G.), NSF DMR 08-46788 (A.L), NSF DMR 09-06780 (P.M.G.), NSF PHY 05-51164 (S.G., A.L.), and Research Corporation through a Cottrell Scholar Award (A.L.).


\begin{thebibliography}{99}
\bibitem{fn1}Although $F$ is insensitive to the quantum statistics, $V$ has contributions from odd as well as even angular-momentum channels in the spin-$\frac{1}{2}$ case. This distinction does not affect the low-energy physics because $V$'s near the fixed point are negligible, compared with $F$'s.
\end{thebibliography}
\end{document}